\documentclass[twocolumn,twocolappendix]{aastex63}
\shorttitle{SMNS survival times}
\shortauthors{Beniamini \& Lu}

\begin{document}
\title{Survival times of supramassive neutron stars resulting from binary neutron star mergers}
\correspondingauthor{Paz Beniamini}
\email{paz.beniamini@gmail.com}
\author{Paz Beniamini}
\affiliation{Theoretical Astrophysics, Walter Burke Institute for Theoretical Physics, Mail Code
350-17, Caltech, Pasadena, CA 91125, USA}
\affiliation{Astrophysics Research Center of the Open University (ARCO), The Open University of Israel, P.O Box 808, Ra’anana 43537, Israel}
\author{Wenbin Lu}
\affiliation{Theoretical Astrophysics, Walter Burke Institute for Theoretical Physics, Mail Code
350-17, Caltech, Pasadena, CA 91125, USA}
\affiliation{Department of Astrophysical Sciences, Princeton University, Princeton, NJ 08544, USA}

\begin{abstract}
A binary neutron star (BNS) merger can lead to various outcomes, from indefinitely stable neutron stars, through supramassive (SMNS) or hypermassive (HMNS) neutron stars supported only temporarily against gravity, to black holes formed promptly after the merger. Up-to-date constraints on the BNS total mass and the neutron star equation of state suggest that a long-lived SMNS may form in $\sim 0.45-0.9$ of BNS mergers. A maximally rotating SMNS needs to lose $\sim 3-6\times 10^{52}$ erg of it's rotational energy before it collapses, on a fraction of the spin-down timescale. A SMNS formation imprints on the electromagnetic counterparts to the BNS merger. However, a comparison with observations reveals tensions. First, the distribution of collapse times is too wide and that of released energies too narrow (and the energy itself too large) to explain the observed distributions of internal X-ray plateaus, invoked as evidence for SMNS-powered energy injection. Secondly, the immense energy injection into the blastwave should lead to extremely bright radio transients which previous studies found to be inconsistent with deep radio observations of short gamma-ray bursts.
Furthermore, we show that upcoming all-sky radio surveys will constrain the extracted energy distribution, independently of a GRB jet formation. Our results can be self-consistently understood, provided that most BNS merger remnants collapse shortly after formation (even if their masses are low enough to allow for SMNS formation).  This naturally occurs if the remnant retains half or less of its initial energy by the time it enters solid body rotation.
\end{abstract}
\keywords{gamma-ray burst: general  -- gravitational waves -- stars: jets -- stars: neutron}


\section{Introduction}
\label{sec:Intro}
The electromagnetic (EM) appearance of binary neutron star (BNS) mergers depend strongly on the nature and evolution of the post-merger remnant, which in turn depends on the component masses as well as the equation of state (EoS) of matter in the deep interior of neutron stars (NSs) \citep{shibata19_bns_merger_review}.
Numerical simulations of the gravitational wave (GW) driven dynamical merger process show that the initial remnant is supported against gravitational collapse by strong differential rotation and partially by thermal pressure \citep{shibata06_merger_simulation, Baiotti2008, sekiguchi11_GW_neutrino_emission,hotokezaka13_merger_remnant, kiuchi14_high_rel_merger_simulation, kaplan14_thermal_pressure}.
Subsequently, the system evolves on the viscous timescale that is roughly $\alpha^{-1}\lesssim 100$ times longer than the rotational period of $\sim 1\rm\,ms$, where the \citet{shakura73_viscous_evolution} dimensionless viscosity parameter is found to be $\alpha\gtrsim 10^{-2}$ in the outer envelope of the remnant NS as given by magneto-hydrodynamic (MHD) turbulence \citep{Kiuchi2018}. However, simulations have not been able to capture the full range of lengthscales and physical processes needed to understand the transport of angular momentum by e.g., Kelvin-Helmholtz and magneto-rotational instabilities \citep[e.g.,][]{kiuchi14_high_rel_merger_simulation, Ciolfi2019}. A simpler approach is to carry out (two-dimensional) viscous hydrodynamic simulations, under the assumption that sub-grid MHD processes operate efficiently to generate a macroscopic viscosity \citep[e.g.,][]{shibata17_2D_viscous_HD, radice17_viscous_HD, fujibayashi18_2D_viscous_HD}. It is found that the outer envelope viscously spreads into a torus, which contains $\sim 0.1M_\odot$ and a large fraction of the angular momentum. However, the evolution of the rotational profile of the neutron star (e.g., how it approaches uniform rotation) and the share of angular momentum between the NS and the torus are highly uncertain as they depend sensitively on the choice of viscosity prescriptions. On longer timescales $\sim 1\rm\,s$, neutrino cooling removes thermal energy ($\sim0.05M_\odot c^2$), increases the central density of the NS, and may cause marginally stable systems to collapse. If at the end of this phase the NS has a sufficiently low mass and sufficiently high angular momentum, it will likely settle into uniform rotation, and become a supramassive neutron star (SMNS), supported against gravity by its fast rotation.

Observations of NSs near two solar masses in a number of sources \citep{demorest10_massive_NS, antoniadis13_massive_NS, cromartie20_massive_NS} suggest that the pressure above nuclear saturation density of $2.8\times10^{14}\rm\, g\,cm^{-3}$ must be sufficiently high such that the maximum mass of non-rotating NSs, $M_{\rm max}$, is greater than $2M_\odot$ \footnote{It has been argued that electromagnetic observations of GW170817 place an upper limit of $M_{\rm max}\lesssim 2.3M_{\odot}$ \citep{MM2017,Granot2017lessons, shibata17_Mmax, rezzolla18_Mmax}.}. Centrifugal support due to uniform rotation allows the maximum mass to be up to about 20\% higher than that of the non-rotating spherical configuration \citep[e.g.,][]{cook94_Mmax_rotating, breu16_Mmax_rotating}. This means that a fraction of BNS merger remnants, at least the ones\footnote{For instance, the PSR J1946+2052 BNS system \citep[total mass $2.5\pm 0.04M_\odot$,][]{stovall18_BNS_lowest_mass}, are expected to produce a SMNS, after accounting for mass loss of a few percent $M_\odot$ or more due to baryonic kilonova ejecta \citep[e.g.,][]{radice18_ejecta_mass} and then $\sim0.1M_\odot$ from GW and neutrinos \citep{hotokezaka13_merger_remnant, bernuzzi16_gw_energy}.} $\lesssim 2.4M_\odot$ (and possibly even higher mass ones), could be spinning sufficiently rapidly as SMNSs, which undergo secular spin-down on longer timescales $\gg 1\rm\, s$. 

Magnetic spin-down from a long-lived ($\gg 1\rm\, s$) SMNS or stable NS provides a powerful source of energy injection, which has an important impact on the EM counterparts of BNS mergers. For instance, the baryonic ejecta may be strongly heated by the non-thermal radiation from the pulsar wind nebulae and produce a UV-optical transient that is much brighter than the traditional radioactive-decay powered kilonova/macronova \citep{kasen10_magnetar_boosted_SN, yu13_merger_nova, MP2014}. The non-thermal radiation in the nebula may escape the ejecta when the bound-free optical depth is less than unity, generating X-ray emission at the level of the spin-down luminosity lasting for a spin-down time or until the SMNS collapses into a black hole (BH) \citep{zhang13_magnetar_x-ray, MP2014,Murase2018}. It has also been proposed that the long-lasting plateau seen in the X-ray lightcurve of some short gamma-ray bursts (GRBs) are produced by the nebula emission \citep{Rowlinson2010,Metzger2011,Dall'Osso2011}. On the other hand, the ejecta acquires a large kinetic energy comparable to the rotational energy of SMNS, and when decelerated by the surrounding medium, it produces bright multi-band afterglow emission \citep{gao13_magnetar_afterglow,MB2014}.

In this paper, we study the distributions of the survival time and the emitted energy from SMNSs prior to the collapse, crucial ingredients for determining the EM counterparts of such systems (see e.g. \citealt{RaviLasky2014}). The goal is to make predictions based on known information (Galactic BNS statistics, LIGO observations, current EoS constraints; \S \ref{sec:Methods}, \ref{sec:Results}) and then compare with observations (\S \ref{sec:Comparison}). We will show that these different components, brought together, reveal a tension. The implications which could provide insights onto the early stages of the merger remnant's evolution and its stability are discussed in \S \ref{sec:conclude}.

\section{Method}
\label{sec:Methods}
As we are interested in systems surviving for time-scales longer than the GRB prompt duration we focus on cold and uniformly rotating NSs, assuming that the differential rotation has subsided (on timescale of $\sim 0.1\rm\, s$) and neutrino cooling has ended (on timescale of $\sim 1\rm\, s$).
We apply realistic equations of state (EOS) using the {\bf rns} code \citep{Stergioulas1995} \footnote{\url{http://www.gravity.phys.uwm.edu/rns/}} to simulate asymmetric models of uniformly rotating cold NSs\footnote{To avoid models that are unstable against radial perturbations we only use models that have a central energy density below a critical value defined by the requirement that it leads to the maximum mass of a non-rotating star \citep{Stergioulas1995}.} and consider different energy loss mechanisms.
The latter is typically dominated by dipole spin-down which in turn depends on the magnetic field strength on the merger remnant's surface.

Our calculation proceeds as follows. We consider first the observed sample of 11 Galactic binary systems with well determined individual gravitational masses (listed in table \ref{tbl:MBNS}) to simulate (gravitational) chirp masses and secondary to primary mass ratios: $M_{\rm ch},q\equiv M_2/M_1\leq 1$ according to observations\footnote{The cosmological distribution of BNS masses may be different between the Galactic population and the cosmological one. For example, it is reasonable to expect that the NS masses could be sensitive to the metallicity, which for cosmological events would span a wider range than for the relatively young Galactic BNS population. However, this will tend to widen the distribution of survival times and therefore strengthen our results discussed below. Taking the Galactic distribution as a proxy is thus a conservative choice for the purposes of this analysis.}. The observed sample can be described by independent distributions. $M_{\rm Ch}$ is fit by a normal distribution with $\mu_{M_{\rm Ch}}=1.175M_{\odot}, \sigma_{M_{\rm Ch}}=0.044M_{\odot}$. The parameter $\tilde{q}=(1-q)/q$ can be fit with an exponential distribution with $\lambda_{\tilde{q}}=0.0954$. This description ensures that $0\leq q \leq 1$.
$M_{\rm ch},q$ are used to calculate $M_1, M_2$ according to
\begin{equation}
	M_1=M_{\rm Ch}q^{-3/5}(1+q)^{1/5}\quad,\quad M_2=qM_1
\end{equation}

While the directly measured quantity for an individual NS is its gravitational mass, it is useful to consider the equivalent baryonic mass, as the total baryonic mass is conserved during the merger. The gravitational masses $M_1, M_2$ are converted to baryonic masses $M_{1,0}, M_{2,0}$ using the assumed EoS, under the assumption of zero spin for the individual NSs (this is consistent with the rotation frequencies of Galactic BNS pulsars, which are well below break-up). During the merger, some of the mass is ejected in the form of dynamical ejecta and disk winds. We use the estimates for the (baryonic) mass of the ejecta, $M_{\rm ej,0}$, as a function of $M_{\rm ch},q$ given by \cite{MM2019,Coughlin2019} based on fits to numerical relativity simulations.
The baryonic mass of the remnant is then simply given by $M_{0}=M_{1,0}+M_{2,0}-M_{\rm ej,0}$.

For a given EoS, we find the maximum baryonic mass of a non-rotating star, $M_{\rm max,0}$ and of a star rotating at the mass shedding limit, $M_{\rm th,0}$. For a given $M_{0}$, there are three possibilities. First, $M_{0}>M_{\rm th,0}$. This leads to either a hypermassive neutron star (HMNS), supported only by differential rotation and / or thermal pressure or, for still higher masses, to a prompt collapse. In either case, the result is a very short survival time (assumed below to be $\sim 0.1$ s). The second possibility is $M_{0}<M_{\rm max,0}$. In this case the NS is infinitely stable. The third case is obtained for $M_{\rm max,0}<M_{0}<M_{\rm th,0}$. This is an interesting case, resulting in a finite survival time that we describe in more detail next.

\begin{figure}
	\centering
	\includegraphics[width = 0.4\textwidth]{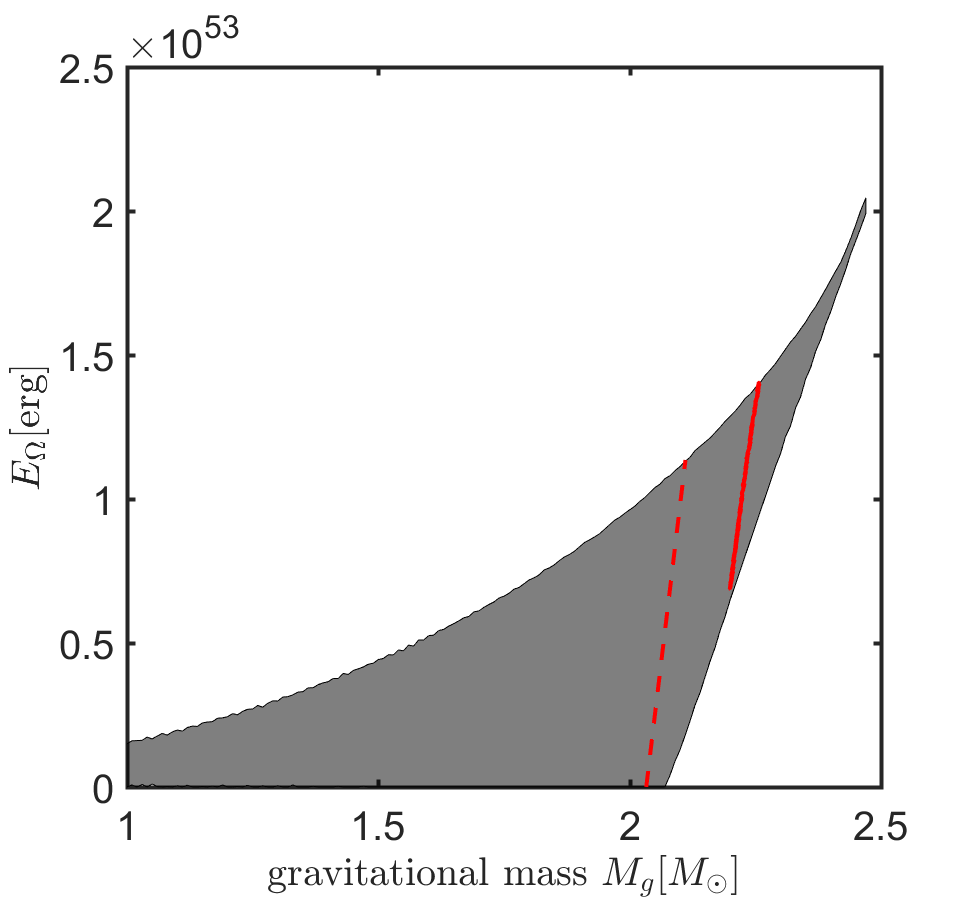}
	\caption{The shaded region depicts allowed solutions (confined by the energy at the mass shedding limit from above and by the minimum required energy for solid rotation from below) for cold uniformly rotating NSs in the plane of gravitational mass and NS rotational energy. The red solid (dashed) line represents a constant baryonic mass of $M_0=2.6M_{\odot}$ ($M_0=2.4M_{\odot}$), corresponding to a gravitational mass of $M_{\rm g}=2.26M_{\odot}$ ($M_{\rm g}=2.11M_{\odot}$) at the mass shedding limit. Results shown here are for the SLy EoS \citep{SLy}.}
	\label{fig:constM0}
\end{figure}

For $M_{\rm max,0}<M_{0}<M_{\rm th,0}$ we construct NS models with the specified EoS such that they have a constant baryonic mass, $M_{0}$ and different rotation rates. We also calculate the maximum extractable energy $E_{\rm ext}=E_{\Omega, \rm max}(M_0)-E_{\Omega, \rm min}(M_0)$ (see also \citealt{Metzger2015}), which is the rotational energy that the NS can lose before it is forced to collapse ($E_{\Omega, \rm max}(M_0)$ is the rotational energy of a maximally rotating NS with baryonic mass $M_0$ and $E_{\Omega, \rm min}(M_0)$ is the minimum rotational energy required for this NS to support itself against collapse). An example of a track with constant baryonic mass in the plane of gravitational mass and rotational energy is shown in Fig. \ref{fig:constM0}. As a limiting case we assume first that the remnant begins maximally rotating (i.e. at the mass shedding limit, \citealt{Giacomazzo2013}). Although possibly unrealistic this is a constructive limit, as lower initial rotation speeds will lead to a quicker collapse and less SMNSs. This second possibility, that the remnant loses a significant amount of its rotational energy while it is differentially rotating is explored in \S \ref{sec:slowrot}, in which we consider an SMNS that enters the cold uniform rotation stage with its rotational energy reduced to $\leq 0.5E_{\Omega, \rm max}(M_0)$. Such a situation is in line with the general-relativistic magnetohydrodynamics (GRMHD) simulations by \cite{Ciolfi2019} which suggest the rotational energy at this stage is $\lesssim 0.1E_{\Omega, \rm max}(M_0)$.

The remnant then spins down according to magnetic dipole radiation, $\dot{E}_{\rm D}$ and gravitational quadrupole radiation, $\dot{E}_{\rm G}$,
\begin{eqnarray}
	\label{Eq:Edot}
	&\dot{E}_{\Omega}=I\Omega \dot{\Omega}+0.5\dot{I}\Omega^2=\dot{E}_{\rm D}+\dot{E}_{\rm G}\\
	& \dot{E}_{\rm D}=-\frac{B_{\rm p}^2R^6\Omega^4}{6c^3}\\
	& \dot{E}_{\rm G}=-\frac{32 I^2 \epsilon^2 G\Omega^6}{5c^5}.
\end{eqnarray}
Where $I$ is the NS moment of inertia, $\Omega$ is its spin frequency, $B_{\rm p}$ is the surface strength of the magnetic field (assumed to be comparable to the poloidal field strength, see \citealt{Reisenegger2001}), we have assumed an aligned rotator for $\dot{E}_{\rm D}$ and $\epsilon$ is the fractional deformation of the NS. The latter can be dominated by different physical effects. One specific deformation mechanism is due to the magnetic field, and given by 
\begin{equation}
	\epsilon=\beta \frac{R^4B_{\phi}^2}{GM^2}
\end{equation}
where $M$ is the gravitational mass of the NS remnant, $B_{\phi}$ is the volume averaged magnetic field (that we have assumed to be dominated by the toroidal field, which can be somewhat greater than $B_{\rm p}$) and $\beta \ll 1$. 
Requiring that the NS interior is stable, puts a limit on the field ratio which can be written as \cite{akgun13_stability_magnetic_configuration}\footnote{We note that there is some uncertainty in the exact limit depending on the assumed structure of the magnetic field and on the equation of state. However a significant increase in the volume averaged $B_{\phi}$ relative to the limit in equation \ref{eq:Bratlim} is needed in order to affect our conclusions below.}
\begin{equation}
	\label{eq:Bratlim}
	\frac{B_{\phi}}{B_{\rm p}}\lesssim 10 \bigg(\frac{M}{2M_{\odot}}\bigg)^{1/2}\bigg(\frac{R}{10\mbox{ km}}\bigg)^{-1}\bigg(\frac{B_{\rm p}}{10^{16}\mbox{ G}}\bigg)^{-1/2}
\end{equation}
Therefore, we take $B_{\phi}=10 B_{\rm p}$ as well as $\beta=0.01$ for our canonical calculation.
For this choice we find that the magnetic dipole radiation dominates the spin-down. Indeed taking into account the stability condition above, we find
\begin{equation}
	\label{eq:GWvsEM}
	\frac{|\dot{E}_{\rm G}|}{|\dot{E}_{\rm D}|} \lesssim \beta^2 I_{45.5}^{2} \bigg(\frac{M}{2.5M_{\odot}}\bigg)^{-2}\bigg(\frac{R}{10\mbox{ km}}\bigg)^{-2}\bigg(\frac{P}{0.6\mbox{ msec}}\bigg)^{-2}
\end{equation}
where $I_{45.5}=\frac{I}{10^{45.5}\mbox{ g cm}^{2}}$. This equation shows that even for $\beta=1$ and rapid initial rotations, stability of the magnetic field configuration implies that GW losses are comparable or sub-dominant to dipole EM losses. We discuss the implications of deviations from these limitations in \S \ref{sec:Results}.

The values of $R,M,I$ depend in general on both the (fixed) $M_0$ and the (evolving) spin rate and are estimated directly from the {\rm rns} models at each time step. $\Omega$ is evolved according to Eq. \ref{Eq:Edot} until the amount of extracted energy equals $E_{\rm ext}$ and the NS collapses. The time elapsed by this point is denoted $t_{\rm survive}$. This calculation is repeated $10^4$ times (each time drawing different parameters from the chirp mass and mass ratio distributions), in order to obtain the distribution of $t_{\rm survive},E_{\rm ext}$ values consistent with Galactic BNS systems.

\begin{table}
	\caption{Galactic binary neutron stars with well determined individual masses. To be consistent with the notation adopted in \S \ref{sec:Methods}, we denote by $M_1$ the more massive of the NSs in the binary. This should not be confused with a standard notation in binary pulsar literature, by which $M_1,M_2$ are used to distinguish the observed pulsar from its companion.}
	\resizebox{0.425\textwidth}{!}{
		\begin{tabular}{cccc}
			\hline 
			System & $M_1 [M_{\odot}]$  & $M_2 [M_{\odot}]$  & reference
			\tabularnewline
			\hline 
			J0737-3039  & 1.338  & 1.249  & \cite{Kramer2006} \tabularnewline
			J1906+0746  &  1.323 & 1.291  & \cite{Lorimer2006} \tabularnewline
			J1756-2251 & 1.341 & 1.23 & \cite{Faulkner2005}  \tabularnewline
			B1913+16 & 1.44 & 1.389 & \cite{Weisberg2010}  \tabularnewline
			B1534+12  & 1.346 & 1.333 & \cite{Stairs2002}  \tabularnewline
			J1829+2456 & 1.306 & 1.299 & \cite{Champion2005}  \tabularnewline
			J1518+4904 & 1.41 & 1.31 & \cite{Janssen2008}  \tabularnewline
			J0453+1559 & 1.559 & 1.174 & \cite{Martinez2015} \tabularnewline
			J1913+1102 & 1.62 & 1.27 & \cite{Lazarus2016} \tabularnewline
			J1757-1854 & 1.3946 & 1.338  & \cite{Cameron2018} \tabularnewline
			J0509+3801 & 1.46 & 1.34  & \cite{Lynch2018} \tabularnewline
			\hline 
			\label{tbl:MBNS}  &  &  & \tabularnewline
	\end{tabular}} 
\end{table}

\section{Results}
\label{sec:Results}
\begin{figure}
	\centering
	\includegraphics[width = 0.4\textwidth]{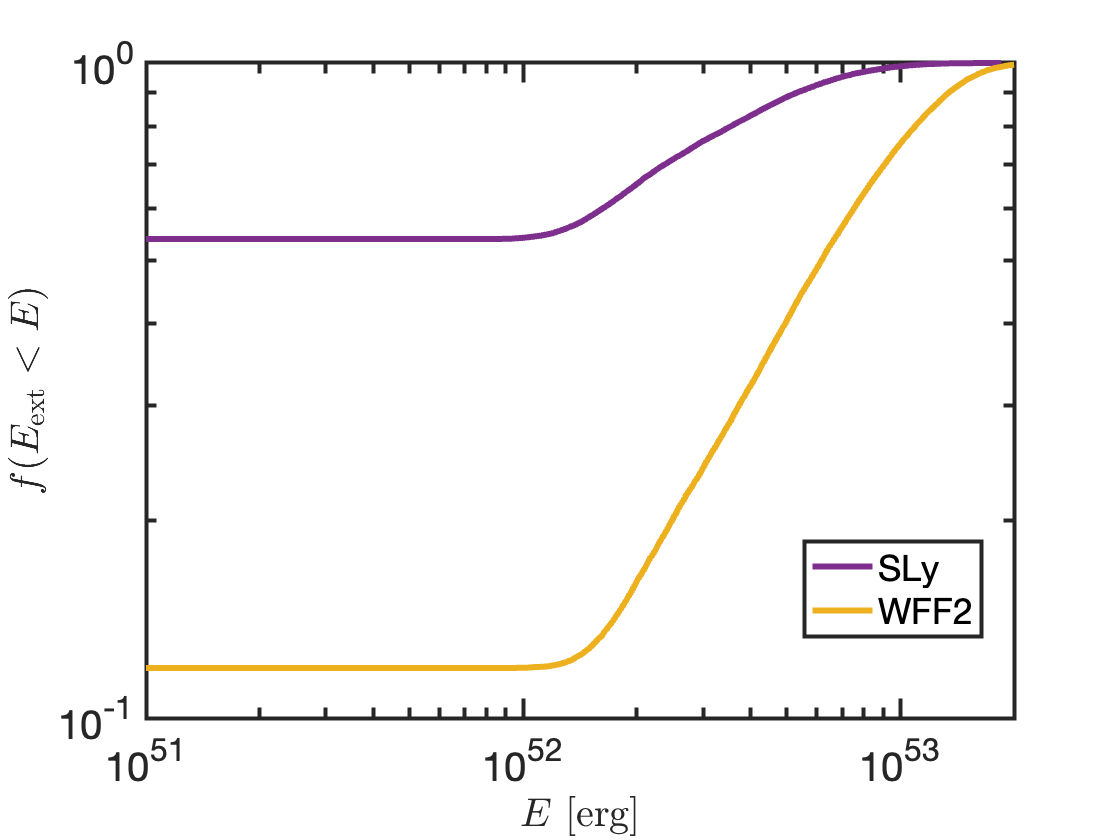}\\
	\includegraphics[width = 0.4\textwidth]{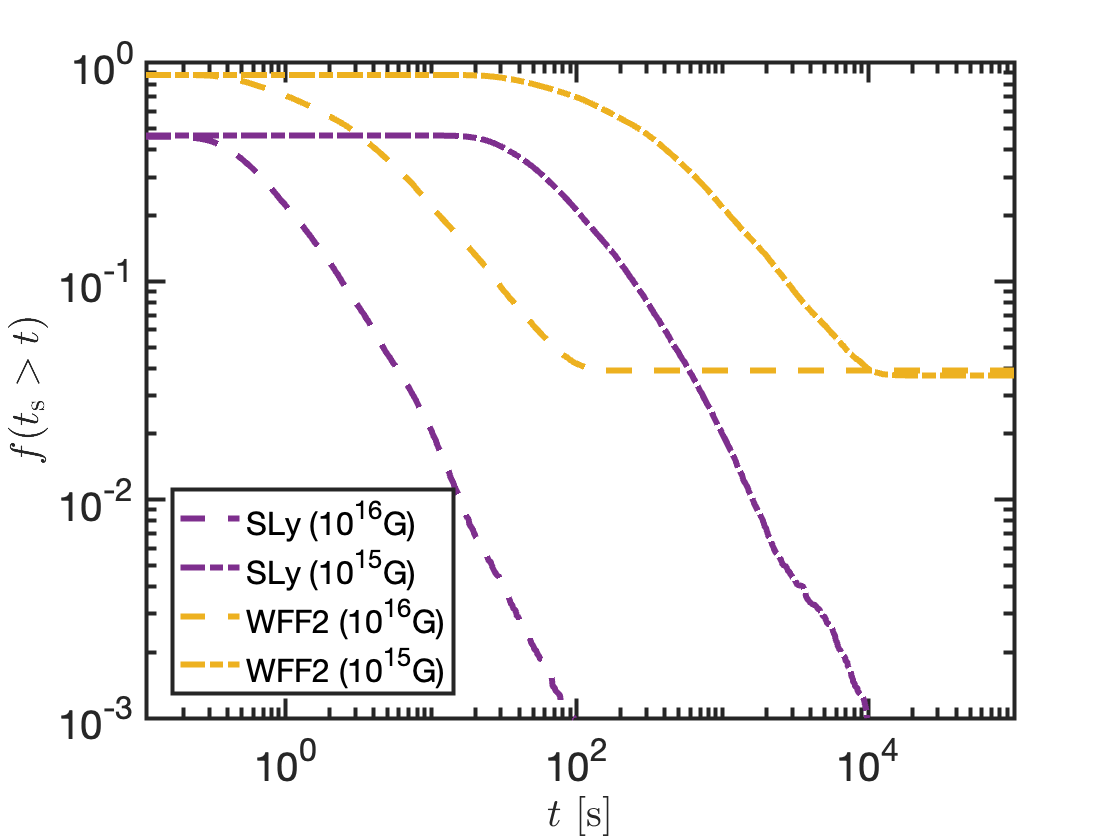}
	\caption{Cumulative distribution of the energies extracted (top) and survival times (bottom) of BNS merger remnants before they collapse to black holes. Results are calculated under the assumption that the merger remnant is born rotating at the mass shedding limit and shown for different EoS and, in the lower panel, for different surface field strengths.}
	\label{fig:tandEdist}
\end{figure}

In figure \ref{fig:tandEdist} we show the distribution of extracted energies and survival times for different EoS and different surface magnetic field strengths.
We consider in particular the SLy EoS \citep{SLy} which is consistent with available observational constraints on the mass-radius curve of NSs \citep{Coughlin2019}. As a comparison case, we consider also the WFF2 EoS \citep{Wiringa1988}. The maximum masses of non-rotating NSs with these EoSs are $M_{\rm max}=2.05$ (SLy) and $M_{\rm max}=2.2$ (WFF2). These EoSs therefore roughly bracket the allowed range of maximum masses allowed by observations. A large fraction ($\sim 45-90\%$) of the merger remnant population have baryonic masses in the range $M_{\rm max,0}<M_{0}<M_{\rm th,0}$ which, in case the remnants begin as maximally rotating, correspond to formation of SMNSs and finite survival times. The median survival time under these assumptions is 80 s (330 s) for $B_{\rm p}=10^{15}$\,G and SLy (WFF2) [0.9 s (3.5 s) for $B_{\rm p}=10^{16}$\,G and SLy (WFF2)]. Approximately these times are $\sim 0.35 \tau_{\rm D,0}$ (where $\tau_{\rm D,0}$ is the initial magnetic dipole spin-down time). This reflects the fact that the NS typically needs to lose $\sim 0.35$ of its rotational energy before collapsing, and for $t<<\tau_{\rm D,0}$ the evolution of the rotational energy is approximately $E_{\Omega}\approx E_{\Omega,0}(1-t/\tau_{\rm D,0})$ (assuming GW quadrupole radiation to be negligible, which is found to be the case in these calculations, see equation \ref{eq:GWvsEM}).

As mentioned above, the merger remnant must lose a large fraction of its initial rotational energy, with a median loss of $\sim 3\times 10^{52}$\, erg for the SLy EoS ($\sim 6\times 10^{52}$\, erg for the WFF2 EoS) before collapsing and with narrow deviations around those values ($\sigma_{\log_{10}E_{\rm ext}}\approx 0.24$ in both cases). This huge amount of energy is much larger than that of the (collimation corrected) GRB jet or that of sub-relativistic ejecta powering the kilonova emission. As a result, a large fraction of this energy could be absorbed by the ejecta and dominate its kinetic energy budget at late times. Such large amounts of kinetic energy can be seen as strong radio emitters on timescales of years to tens of years, as addressed in more detail in \S \ref{sec:magboost}. 

Interestingly, even for a fixed magnetic field strength across different systems, the distribution of $t_{\rm survive}$ for SMNSs is very wide, with $\sigma_{\log_{10}t_{\rm survive}}\approx 0.5$. Any spread in the surface field strength of the merger remnant (which is very natural), would only cause the distribution of $t_{\rm survive}$ to be much wider still. We return to this point in \S \ref{sec:plateaus}.

If $B_{\phi}/B_{\rm p}$ and/or $\beta$ are much greater than our canonical choices described in \S \ref{sec:Methods}, GW radiation may become significant enough to dominate the spin-down (note that as shown in equation \ref{eq:GWvsEM} this may lead to an unstable magnetic field configuration). The time to collapse in this situation will be shorter than without GW emission. Nonetheless, the collapse time distribution (in log space) is wider than with dipole spin-down only. This is because, even for a single magnetic field value, there is a range of masses resulting in a SMNS, each corresponding to a different energy that should be lost before collapse. The distribution of extracted energies prior to collapse, depends sensitively on $M_{\rm th}\!-\!M$ and is independent of the energy loss mechanism. As the energy that needs to be removed before collapse is typically a fraction of the initial rotational energy, the collapse time is typically a fraction of the spin-down time. For GW emission, the latter has a stronger dependence on the initial period as compared with the dipole case ($\tau_{\rm G,0}\propto P^{4}$ rather than $\tau_{\rm D,0}\propto P^2$) and may also have a dependence on the mass (depending on what sets the value of $\epsilon$, see \S \ref{sec:Methods}). This means that even for all other parameters being fixed, the variation in the spin-down time and time to collapse will be comparable (and in fact, slightly greater) in the case where losses are dominated by GW emission.
Furthermore, in case the two energy loss mechanisms both play a significant role in spin-down, their spin-down timescales depend differently on the various parameters, causing a further widening of the collapse time distribution. 
In that aspect, the analysis including dipole losses only, is conservative.

\subsection{Slower initial rotations and fallback accretion}
\label{sec:slowrot}
As mentioned in \S \ref{sec:Methods}, GRMHD simulations point towards the possibility that the merger remnant losses a significant fraction of its initial rotation energy during the differential rotation phase \citep{Kiuchi2018,Ciolfi2019}. The energy lost at this phase is mostly converted to internal energy of the remnant. Following the results of \cite{Ciolfi2019}, we consider here the possibility that the merger remnant enters the cold uniform rotation phase with a rotational energy of $\leq 0.5E_{\Omega, \rm max}(M_0)$. We note that the physical processes responsible for the angular momentum transport in the differential rotation phase are still highly uncertain. \citet{radice18_remnant_angular_momentum} argued that the total angular momentum of the merger remnant often exceeds the mass-shedding limit for SMNS (favoring the $E_{\Omega}=E_{\Omega,\rm max}$ prescription considered earlier), but their simulations did not include an explicit treatment of angular momentum transport after the GW-driven dynamical merger phase.

\begin{figure}
	\centering
	\includegraphics[width = 0.4\textwidth]{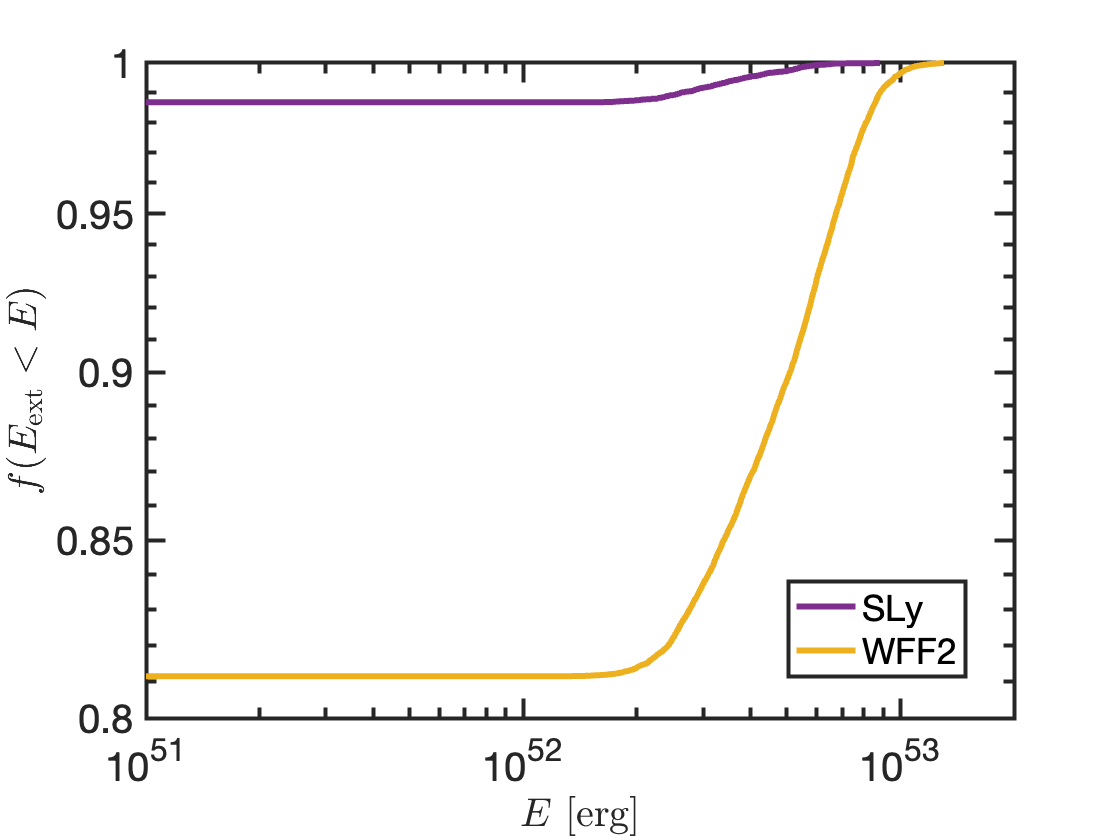}\\
	\includegraphics[width = 0.4\textwidth]{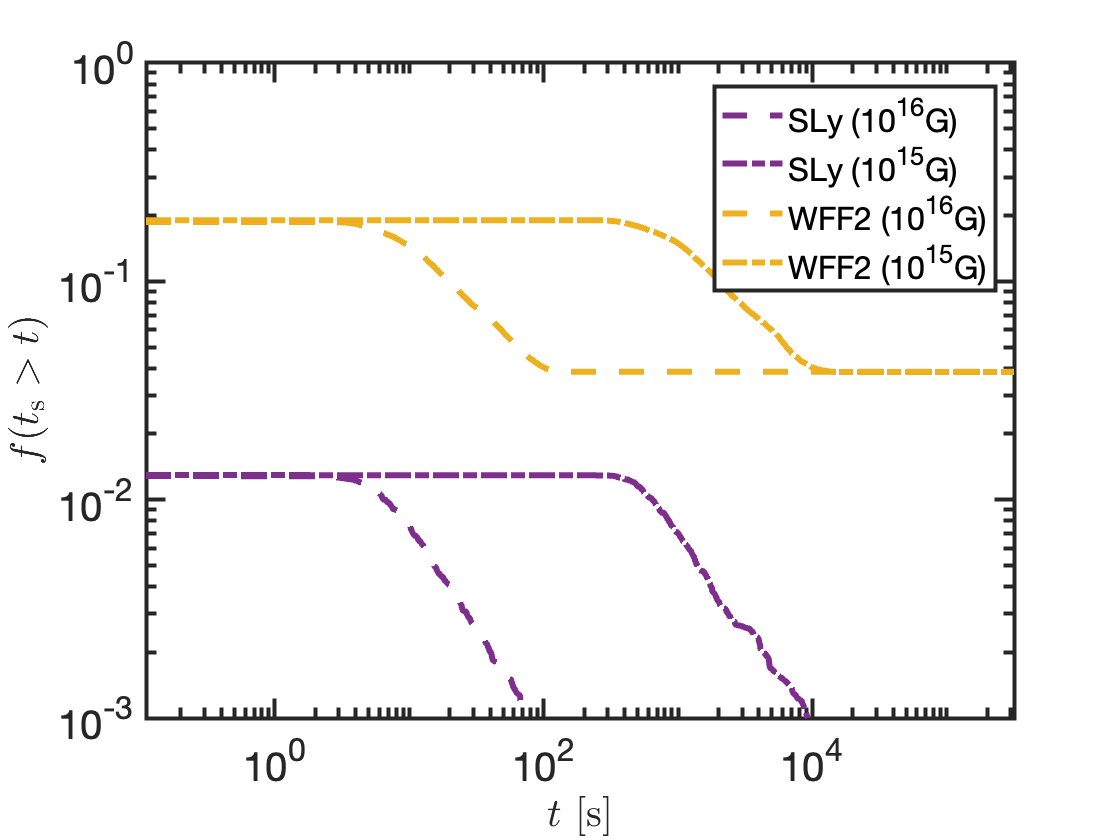}
	\caption{Same as Figure \ref{fig:tandEdist}, but assuming that the NS merger remnant enters the cold uniform rotation phase with a rotational energy equaling $0.5E_{\Omega, \rm max}(M_0)$ (note the different scale for the energy plot as compared to Figure \ref{fig:tandEdist}).}
	\label{fig:tandEdisthalf}
\end{figure}

We show in figure \ref{fig:tandEdisthalf} the distributions of energies extracted and survival times prior to collapse for this case of slower initial rotation. The main difference compared to the case in which the merger remnant is born rotating at the mass shedding limit, is the fraction of systems that end up as SMNSs. This fraction is $1.5\%$ ($15\%$) for the SLy (WFF2) EoS. The implication is that, depending on the EoS, considering realistic estimates for the angular momentum loss during the differential rotation phase it is possible that only a very small fraction, and perhaps even none of the merger remnants, result in long lived NSs. Indeed if the remnant enters the solid body rotation with even lower energy, for example with $0.1E_{\Omega, \rm max}(M_0)$, then only $0.07\%$ ($3\%$) of systems for the SLy (WFF2) EoS, become long lived remnants. Furthermore, these long lived remnants would only lose a small amount of rotational energies, $\lesssim 10^{52}\mbox{ erg}$, before they collapse.

The survival time and extracted energy distributions can also be modified in the presence of significant fallback accretion onto the newly born strongly magnetized NS --- or magnetar \citep{MBG2018}. In particular, the magnetar may transfer some of its angular momentum to the fallback disk, and as a result, spin down to the point of collapse after having released a smaller amount of its initial rotational energy. \cite{MBG2018} have found that this can decrease the amount of energy released prior to collapse by up to a factor of a few. Similar to the case of slow initial rotation, this will lead to a faster collapse relative to the case of fast initial rotation with no fallback accretion.

\section{Comparison with sGRBs and implications}
\label{sec:Comparison}
\subsection{Can magnetars power the observed plateaus?}
\label{sec:plateaus}
X-ray afterglows of sGRBs sometimes exhibit a `plateau' phase, where the emission is almost steady or declining very slowly with time. Previous studies \citep{ZM2001,Rowlinson2010,Metzger2011,Dall'Osso2011,Rezzolla2015,Lu2015,Stratta2018} have interpreted these as possible evidence of energy injection into the GRB jet or into the surrounding post-merger medium with a duration comparable to the observed plateau (although see \citealt{EG2006,Granot2006,Ioka2006,Genet2007,Shen2012,BDDM2020,Oganesyan2020} for other interpretations). These studies pointed to a magnetar's spin-down as a possible source of this energy injection. Our results shed light on the viability of this interpretation.

Two types of plateaus are observed in sGRBs. The first are `external plateaus'. These are plateaus in which the light-curve smoothly and gradually transitions from a flat temporal evolution to a declining one. More quantitatively, these are cases where the plateau declines on a timescale $\Delta t$ such that $\Delta t/t\sim 1$ (where $t$ is the time since the GRB trigger of the end of the plateau).
These plateaus are dubbed external, since their low level of variability can be reproduced by emission from the external shock (the same region from which the standard afterglow signal is observed). As such it is not possible to separate by these lightcurves between a scenario where energy injection ended abruptly but the energy was then reprocessed at the external shock to a situation where the energy injection itself was simply slow to fade at the end of the plateau phase. Furthermore, as mentioned above, such plateaus do not necessarily require any energy injection whatsoever.

The second type of plateaus are `internal plateaus' in which the emission quickly declines at the end of the plateau (such that  $\Delta t/t\ll 1$ with the same definitions as above). In these cases, the rapid variability strongly suggests energy injection as well as an emission radius well below the external shock. In the context of magnetar energy injection, the source of the abrupt cut-off is most naturally associated with the collapse of an unstable magnetar to a black hole. 

\begin{figure}
	\centering
	\includegraphics[width = 0.5\textwidth]{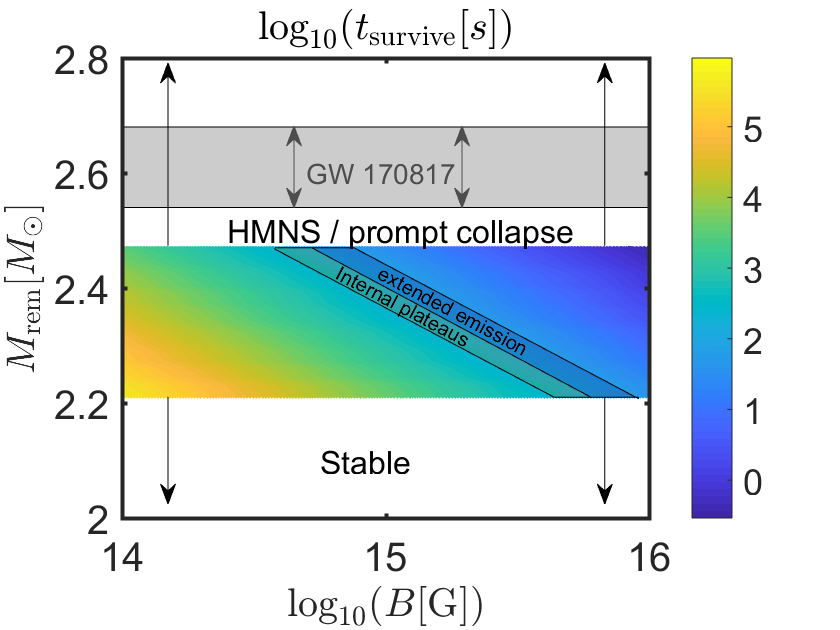}
	\caption{Survival time of a magnetar NS merger remnant as a function of the dipole field strength, $B_{\rm p}$, and the gravitational mass of the remnant as it enters the phase of cold uniform rotation, $M_{\rm rem}$ (this is larger than the final gravitational mass, by the time the remnant has spun down). At large masses, the magnetar promptly collapses to a black hole, while at lower masses the magnetar is indefinitely stable. Finite survival times are possible in between and are very sensitive to both $B_{\rm p}$ and $M_{\rm rem}$. As a comparison, the durations of all SGRBs with `internal plateaus' span a narrow strip in this parameter space, denoted by a diagonal shaded region. Also shown in comparison is the remnant mass of GW170817, which favours a rapid collapse. We assume the remnant lost $0.1-0.2M_{\odot}$ due to GW emission, neutrinos and baryonic ejecta during the merger. Results in this plot are calculated according to the procedure described in \S \ref{sec:Methods} and for the SLy EoS.
	}
	\label{fig:BM2D}
\end{figure}

\cite{Gompertz2020} study a sample of {\it Swift} detected sGRBs with known redshifts. Only 5/22 of the bursts in the sample with an inferred $>1/2$ probability of being non-collapsars (or 9/39 of their full sample), exhibited an internal plateau. This is low, as compared with the fraction $\approx0.45-0.9$ of long lived SMNSs expected to result from neutron star mergers found above. Furthermore, the (source frame) durations of these internal plateaus are tightly clustered, with $<\log_{10}t_{\rm IXP}[\mbox{s}]>=2.15, \sigma_{\log_{10}(t_{\rm IXP})}=0.16$. As mentioned in \S \ref{sec:Results}, this is in contradiction with even the most conservative estimates for the deviation in $t_{\rm survive}$, which is expected to span a wide range of survival times (this is true also in the case where the merger remnant enters the uniform rotation phase with a slower angular velocity and in the case where it undergoes significant fallback accretion early on). This is demonstrated in figure \ref{fig:BM2D}, where we show the survival time as a function of the remnant's dipole field strength and gravitational mass (measured immediately after collapse), under the assumption that the merger remnant is initially rotating at the mass shedding limit. The resulting survival times, span over five orders of magnitude, while the durations of internal plateaus all reside in a narrow strip within this plane. This makes the association of these plateaus with spin-down of SMNS extremely fine tuned. It is also worthwhile to stress that sGRBs with internal plateaus span a range of prompt gamma-ray energies, durations and redshifts consistent with the rest of the sGRB population (see \citealt{Gompertz2020}), suggesting that observational selection effects are unlikely to play a significant role in resolving this apparent contradiction. Finally, it is interesting to consider also the total energies released in the X-ray band during the observed IXPs. We find $<\log_{10}E_{\rm IXP,iso}[\mbox{erg}]>=49.5, \sigma_{\log_{10}(E_{\rm IXP})}=0.9$.
Taking into account the typical beaming of sGRBs (with a jet opening angle of $\theta_0\approx 0.1$, see \citealt{Beniamini2019,Nakar2020}), the collimation corrected energies corresponding to the values above are approximately two orders of magnitude lower. We note however that this correction does not apply if the X-rays are produced by a roughly isotropic component injected by the merger remnant, see \cite{MP2014}.
All together, the energies observed in IXPs are approximately three to five orders of magnitude below the typical energy release expected from a SMNS before collapse (see Fig. \ref{fig:tandEdist}). 
Furthermore, the large spread in plateau energies is also not expected in this scenario (in contrast with the survival time, the energy release is independent of the magnetic field strength and is expected to be quite narrowly distributed). We conclude that the observed plateau data are not naturally explained by energy injection from a SMNS merger remnant.

An additional emission feature seen in $\sim 15-20\%$ of sGRBs is the `extended emission' (EE; \citealt{Lazzati2001,Gehrels2006,Norris2010}), a prolonged feature of soft $\gamma$-rays lasting $\sim 100$\, s after the initial prompt hard spike. Similar to IXPs, the EE terminates on a timescale that is very short relative to their overall duration (and indeed exhibit significant variability during their activity) and may therefore be an indication of SMNS collapse. 
7/39 GRBs in the \cite{Gompertz2020} sample exhibit EE. Their (source frame) durations are $<\log_{10}t_{\rm EE}[\mbox{s}]>=1.85, \sigma_{\log_{10}(t_{\rm EE})}=0.16$. As for the IXPs, this distribution is much narrower than naturally expected from SMNS collapse. Furthermore, if such an association is true, EE sGRBs should result in extremely bright radio remnants (as will be discussed \S \ref{sec:magboost}). \cite{Fong2016} have conducted a specific search in radio for EE sGRBs and their search yielded only upper limits. In particular, for the EE sGRB, 050724, the ejecta energy is constrained to be $<10^{52}$\, erg, much below the energies associated with long lived NS remnants. In addition, the radio search of \cite{Ricci2021} limits the released energy of three of the other EE sGRBs in the \cite{Gompertz2020} sample: $E_{060614}<5\times 10^{52}$\, erg, $E_{061006}<4.5\times 10^{52}$\, erg, $E_{150424A}<3\times 10^{52}$\, erg.

Finally, if the merger remnants enter the solid body rotation phase with significantly slower rotations than the mass shedding limit, then only a small fraction of remnants produce long-lived systems. For example, if they retain $\leq 10\%$ of their initial energy \citep{Ciolfi2019}, then $<3\%$ of BNS mergers lead to long lived remnants (see \S \ref{sec:slowrot}), much less than the occurrence rate of internal plateaus and EE sGRBs. In such a situation the bulk of plateaus / EEs cannot be explained by collapses of SMNS remnants.

\subsection{EM signatures of magnetar boosted outflows}
\label{sec:magboost}
Energy ejected by the magnetar prior to the point of collapse will eventually catch up with the (roughly isotropic) ejecta expanding away from the merger remnant. This energy could then re-energize the ejecta \footnote{The extent to which this happens depends on how efficiently the injected energy is absorbed by the ejecta (see e.g. \citealt{MP2014})} and may significantly overwhelm the initial kinetic energy of the ejecta.

Depending on the timescale of energy injection relative to the diffusion time, this process may lead to a magnetar-boosted kilonova \citep{kasen10_magnetar_boosted_SN, yu13_merger_nova, MP2014}. However, the effect of this energy injection on the radio light-curve is more generic. The peak of the radio lightcurve from the merger ejecta occurs on a timescale of years to tens of years, when the mildly relativistic ejecta has been decelerated by the external medium \citep{NakarPiran2011,Piran2013,PK2013,Hotokezaka2018,Radice2018,Kathirgamaraju2019}. This timescale is longer than $t_{\rm survive}$ for any reasonable value of the magnetic field (that could account for the production of a GRB in the earlier stages). Furthermore, the observed signal depends on the (narrowly distributed) kinetic energy of the ejecta and is independent of the specific time at which energy was added to the ejecta. As a result the radio counterpart is rather robust to the uncertainties in the merger outcome. The flux and duration of the peak are given by
\begin{eqnarray}
	\label{eq:Fpktpk}
	& F_{\nu,\rm pk}= 200 \epsilon_{\rm e,-1}^{p-1} \epsilon_{\rm B,-3}^{p+1\over 4} n_0^{p+1\over 4} M_{\rm ej,-1}^{7-5p \over 4}E_{51}^{5p-3 \over 4} \nu_9^{1-p\over 2}d_{26}^{-2} \mu\mbox{Jy} \\
	& t_{\rm pk}=3.5 n_0^{-1/3} E_{51}^{-1/2} M_{\rm ej,-1}^{5/6}\mbox{ yr}
\end{eqnarray}
where $E$ is the energy of the blastwave, $n$ is the external density, $d$ is the distance of the explosion, $\epsilon_{\rm e},\epsilon_{\rm B}$ are the fractions of the shock energy, deposited in relativistic electrons / magnetic fields respectively, $p$ is the index of the shocked electrons' PL energy distribution and $M_{\rm ej}$ is the total ejecta mass.
The numerical coefficients in Eq. \ref{eq:Fpktpk} depend in general on $p,\alpha$ (where $\alpha$ describes the distribution of velocities in the ejecta, $E(>\beta\Gamma)\propto (\beta\Gamma)^{-\alpha}$) and are estimated here for $p\approx2.2, \alpha\approx 4$ (e.g. \citealt{NakarPiran2017,Kathirgamaraju2019}). This choice of $\alpha$ corresponds to a steep energy profile, meaning that the results are weakly dependent on its exact value, and deriving the same expressions assuming only a single velocity to the ejecta would only have introduced an order unity change.
We also employ the convention $q_X\equiv q/10^X$ in cgs units (except for $M_{\rm ej}$ which is in units of $M_{\odot}$).
With the large values of extracted energies given in \S \ref{sec:Results}, we can expect a large number of such bright radio transients.

\begin{figure}
	\centering
	\includegraphics[width = 0.45\textwidth]{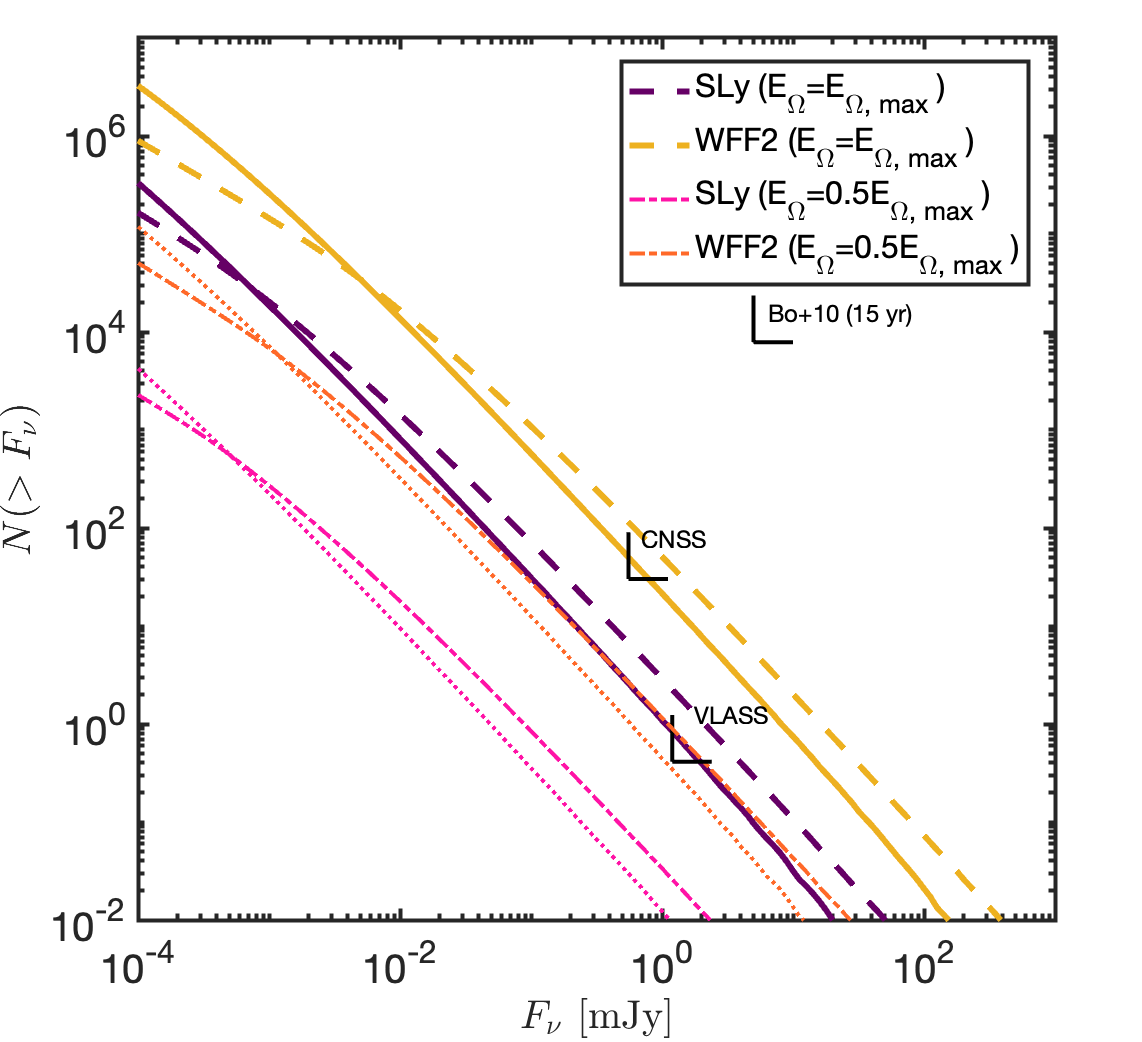}
	\caption{All-sky number of radio sources above a given flux threshold. Results are shown for observations at 3 GHz and for $\epsilon_e=10^{-1},\epsilon_B=10^{-3}$. Solid (dotted) lines depict an external density $n=10^{-2}\mbox{ cm}^{-3}$ for the entire population, while dashed (dot-dashed) lines show the potential contribution assuming a small fraction of the BNS mergers, taken here as $5\%$, occur at an external density of $n=1\mbox{ cm}^{-3}$. The extracted energy and ejecta mass distributions correspond to the two EoSs explored in this work, assuming that the remnant enters the phase of cold uniform rotation with a rotational energy at the mass shedding limit (and half that for dotted, dashed-dotted curves). For comparison we show sensitivity limits of different (planned) surveys \citep{Bower2010,Mooley2016,Lacy2020}.}
	\label{fig:fluxdist}
\end{figure}

Previous studies have considered radio follow-up observations of known sGRBs years after the bursts, and limits have been put on the kinetic energies of the ejecta in those bursts \citep{MB2014,Fong2016,Horesh2016,Ricci2021}. \cite{Ricci2021} found that for ejecta masses $M_{\rm ej}<10^{-2}M_{\odot}$ ($M_{\rm ej}<5\times 10^{-2}M_{\odot}$), extracted energies $>5\times 10^{52}$erg ($> 10^{53}$erg) can be ruled out in all of the 17 bursts studied. If sGRBs require a black hole central engine, then it is not surprising that the ejecta kinetic energy is much below $3\times10^{52}\rm\, erg$. However, if there is a long-lived SMNS in a large fraction of sGRBs, we would have expect larger energies (see figure \ref{fig:fluxdist}).

An independent approach would be to look for bright radio transients in a blind survey \citep{MBW2015}. We calculate the number of sources (all-sky) above a given flux, resulting from mergers with different EoS and with extracted energy and ejecta mass distributions following our results in \S \ref{sec:Results} as
\begin{equation}
	N(>F_{\nu})=\int \int \frac{d\mathcal{R}(z)}{dM_{\rm ej}}\frac{ t(F_{\nu}|M_{\rm ej},z)}{1+z}\frac{dV}{dz} dM_{\rm ej}dz.
\end{equation}
where $d\mathcal{R}(z)/dM_{\rm ej}$ is the co-moving rate of mergers with ejecta mass $M_{\rm ej}$ and $t(F_{\nu}|M_{\rm ej},z)$ is the duration over which a merger with ejecta mass $M_{\rm ej}$ and from redshift $z$ will reside above a flux $F_{\nu}$. The factor of $1+z$ in the denominator is in order to convert from the co-moving to the observed rates and $dV/dz$ is the change in a co-moving volume element with the redshift.
The redshift dependence of the merger rate is obtained by convolving the cosmic star-formation rate \citep{MD2014} with the delay time distribution between binary formation and merger, found from Galactic binary neutron stars \citep{BP2019}. The normalization for the rate of mergers is taken as $\mathcal{R}=320\mbox{ Gpc}^{-3}\mbox{ yr}^{-1}$, in line with the most recent constraints from LIGO-VIRGO \citep{LIGO2020}. Finally, we have taken advantage of the fact that for a given EoS, the extracted energy is to a good approximation dictated uniquely by the ejecta mass (roughly $E\propto M_{\rm ej}^{3.2}$ for SLy and $E\propto M_{\rm ej}^{1.2}$ for WFF2 for remnants that are initially rotating at the mass shedding limit, see appendix \ref{app:MejEext}). In both cases the scaling holds above a minimum $M_{\rm ej}$ for which the production of a SMNS is possible, see \S \ref{sec:Results}).

In Fig. \ref{fig:fluxdist} we plot the results for the all-sky rates. We assume the merger remnant to be initially spinning at the mass shedding limit. We also take relatively conservative choices for the values of the microphysical parameters, $\epsilon_{e,-1}=\epsilon_{B,-3}=1$ and the external density, $n_0=10^{-2}$. We note that since sGRBs result from BNS mergers (which themselves are delayed relative to star formation), they occur in lower external densities than lGRBs. Nonetheless, \cite{O'Connor2020} have recently studied sGRB afterglows and found that the majority of sGRBs take place at environments with $n\gtrsim 10^{-2.5}\mbox{ cm}^{-3}$ (consistent also with the relatively short delay times between binary formation and merger inferred from Galactic BNSs, \citealt{BP2019}) and some fraction at even much greater densities. With this in mind, we consider also the contribution from a $5\%$ sub-population of BNS mergers that take place at greater external densities $n\approx 1\mbox{ cm}^{-3}$. At the high flux end, the curves in Fig. \ref{fig:fluxdist} follow to a good approximation the scaling $N(>F_{\nu})\propto F^{-3/2}$ expected for standard candles in Euclidean geometry. We compare the resulting distributions with the planned sensitivities of the PiGSS, CNSS and VLASS radio surveys \citep{Bower2010,Mooley2016,Lacy2020}. Our results suggest that if BNS mergers enter the cold uniform rotation phase with a rotational energy close to that of the mass shedding limit, then their radio counterparts should be detectable by the VLASS and potentially also the CNSS surveys.
To demonstrate the importance of the initial energy of the remnant as it enters the cold uniform rotation phase, we also plot in Fig. \ref{fig:fluxdist} the results for the case in which the remnant enters the solid body rotation with $E_{\Omega}=0.5E_{\Omega, \rm max}(M_0)$. Due to the lower extracted energies and the overall smaller fraction of long lived remnants, this modest change in the initial energy corresponds to a reduction of $N(>F_{\nu})$ by a factor of $\sim 30-60$. Reducing the initial energy even further, $E_{\Omega}=0.1E_{\Omega, \rm max}(M_0)$, corresponds to a decrease in $N(>F_{\nu})$ by a factor of 5000 compared to the results for the mass shedding limit.
Clearly, constraining the all-sky number of radio sources is a promising avenue towards testing the nature of the remnants of BNS mergers.

At times greater than $t_{\rm pk}$, the kilonova ejecta's velocity becomes dependent only on the blastwave energy and not on its mass. Thus, once the ejecta slows down to Newtonian velocities, the emission becomes very similar to that of a supernova remnant. On the one hand, this is a drawback, as simply by virtue of their rates, there are $\sim 10^3$ times more ``garden variety" supernova remnants than there are kilonova (or magnetar boosted kilonova) remnants. However, these older kilonova remnants are also several orders of magnitude more common than the years old kilonova afterglow transients mentioned above. This means that there could be many such sources even in our own Galaxy or in the local group (the latter may be preferable, as the remnants will cover a smaller area of the sky and their distance can be well determined independently of their angular size). This could be extremely constraining regarding the fate of BNS mergers, provided that these remnants can be reliably identified. A full exploration of this idea is deferred to a future work. That being said, we briefly illustrate here the viability of such an endeavour. Consider a remnant that is $t=10^5$\,yrs old. Given the Galactic BNS merger rate of $\sim 30\mbox{ Myr}^{-1}$ \citep{HBP2018} and the mass of M31, relative to our Galaxy, there should be several such sources in M31. Assuming an overall energy in the blastwave of $\sim 3\times 10^{52}\mbox{ erg}$ (typical of a SMNS remnant, see \S \ref{sec:Results}) and an ISM density of $0.1\mbox{ cm}^{-3}$, these sources should be in the Sedov Taylor phase (rather than the latter ``radiative" phase, see, e.g. \citealt{BWG2016} and references therein). The integrated synchrotron flux can then be calculated according to the ``deep Newtonian" formulation \citep{SironiGiannios2013}. At 1 GHz, and taking $\epsilon_{\rm e}=0.1,\epsilon_{\rm B}=0.01$, we find $F_{\nu}=5 \mbox{ mJy}$. The radius of this remnant is approximately 100 pc, corresponding to an angular size of $\sim 0.3$\, arcmin. The surface brightness is therefore $12 \mbox{ mJy arcmin}^{-2}$, making it potentially detectable by several existing  and planned radio surveys such as NVSS, SUMSS, WODAN and EMU \citep{Intema2017}. The main challenge of such a search would be to reliably infer the energy, independently of $n, \epsilon_{\rm e}, \epsilon_{\rm B}$. If this can be done, then a remnant with an estimated energy $\gtrsim 3\times 10^{52}\mbox{ erg}$ would be a ``smoking gun" evidence of a BNS remnant that produced a long lived magnetar. Conversely, if the existence of such energetic remnants can be ruled out, it would favour remnants that collapse early on to form black holes. 

\section{Conclusions and discussion}
\label{sec:conclude}
We have explored in this work the expected outcome of binary NS-NS mergers as informed by the Galactic BNS population, numerical relativity merger simulations and current constraints on the NS EoS. We assume that a NS merger remnant can cool down and lose differential rotation within the first $\sim 0.3-1$\, s after the merger. Using models of cold and uniformly rotating NSs we calculate the evolution of the merger remnants after this initial phase (we return to the validity of this assumption below). We find that a significant fraction of mergers (0.45 -- 0.9) are expected to end up as SMNS which would survive for finite times before collapsing. The survival time is typically dominated by the dipole spin-down time (the mean survival time is roughly $35\%$ of the spin-down time, corresponding to the mean fraction of rotational energy that needs to be lost before the NS collapses) and its value strongly depends on the final remnant mass and dipole field strength. Even for a fixed dipole field strength between merger remnants, the scatter in the survival time distributions is rather large, with $\sigma_{\log_{10}t_{\rm survive}}=0.5$. If a SMNS is formed, a large amount of energy, $\sim 10^{52}-10^{53}$erg, is extracted from the NS before it becomes unstable and undergoes collapse.

This abruptly terminating energy source has previously been invoked to explain IXPs (internal X-ray plateaus - in which the flux stays roughly constant and then declines very rapidly) seen in short GRB afterglow lightcurves. However, we find that the distribution of observed IXP durations is very narrow while the distribution of released energy in IXPs is very wide. Both behaviours are the opposite to what would be expected from a SMNS collapse and the comparison of the model with observations strongly disfavours the interpretation of energy injection from a pre-collapse SMNS.

The huge amount of energy injected to the blastwave before the SMNS collapses, leads to extremely bright radio transients (due to deceleration of the sub-relativistic ejecta by the surrounding environment). Such sources have already previously been ruled out in all 17 short GRBs in which deep searches have been carried out \citep{Ricci2021}. Furthermore, we calculate here the all-sky rate of such sources and find that at 3 GHz, $1\lesssim N(>1\rm\, mJy)\lesssim 50$ sources are expected. This can be put to the test with existing and future radio surveys. The advantage of this technique over follow-ups of specific GRBs, is that it can constrain also the possibility that the mergers that lead to SMNSs (that inject a large amount of energy to a sub-relativistic outflow) are preferentially those that do not lead to GRBs. Such a trend is reasonable considering that simulations of BNS mergers find unfavourable conditions for the formation of an ultra-relativistic jet (as required for powering GRBs) in cases where a long lived NS is formed \citep{Ciolfi2019,Ciolfi2020}.

An independent line of reasoning stems from the typical time intervals between the binary merger and the launch of a sGRB jet. \cite{BBPG2020} have studied sGRBs with known redshift and shown that the jet-launching delay is typically $\lesssim 0.1$\, s. These results are inconsistent with expectations from the model where sGRBs are powered by long-lived rapidly rotating NSs instead of black hole accretion. The reason is that shortly after its formation, the environment surrounding the magnetar is very baryon rich, preventing the formation of an ultra-relativistic jet, as required for powering the GRB prompt emission (see \citealt{BGM2017} for details). For values of the dipole field consistent with powering short GRBs ($\sim 10^{15}-10^{16}$\,G) the pulsar wind achieve high magnetization (and hence high Lorentz factor) only after a delay of $\gtrsim 10$\, s, which is significantly longer than the values inferred from observed sGRBs ($\lesssim 0.1$\, s, as mentioned above).

All three lines of evidence (IXP, radio afterglow, and jet launching time) shed serious doubts on the formation of long-lived or indefinitely stable magnetars from BNS mergers. This, however, appears to be inconsistent with the estimates of large fractions of such outcomes mentioned at the head of this section. We propose that all these pieces of information can be consistently resolved if merger remnants tend to collapse early on after the merger --- while the proto-NS is still undergoing differential rotation and/or neutrino cooling. This can be tested in the future with detailed GRMHD simulations of differentially rotating NSs with neutrino cooling. Although such a calculation is beyond the scope of this work, we roughly outline below why such an outcome is at least plausible.

After GW emission becomes unimportant, the remnant has a slowly rotating core and rapidly rotating envelope \citep[e.g.,][]{kiuchi14_high_rel_merger_simulation, hanauske17_rotational_profile, Ciolfi2019}. The angular frequency $\Omega$ increases with circumferential radius $r$ until a maximum is reached, and then the rotational rate gradually drops and asymptotically approaches the Keplerian rate at large radii. Such a differentially rotating system can be divided into two regions: (1) in the outer region where $\mathrm{d}\Omega/\mathrm{d} r < 0$, the magneto-rotational instability \citep[MRI,][]{balbus98_MRI} generates strong MHD turbulence/dissipation and hence leads to efficient transport of angular momentum; (2) in the inner region where $\mathrm{d} \Omega/\mathrm{d} r > 0$, MRI does not operate, but the free energy associated with differential rotation is spent to amplify the toroidal magnetic field which grows and saturates due to non-linear dissipation --- such energy dissipation tends to push the system towards uniform rotation, because the system has the general tendency of evolving towards the minimum energy state \citep{lyndenbell74_accretion_disk}. However, the mechanism for magnetic energy dissipation as well as the dissipation rate is currently unknown (the dissipation in above-mentioned simulations are largely due to numerical viscosity).

The magnetic energy dissipation timescale may be written as some multiplicity factor $\xi>1$ times the Alfv{\'e}n crossing time $t_{\rm A}\sim \sqrt{4\pi \rho R^2}/B_\phi\sim 10\mathrm{\,ms}\, B_{\phi,16}^{-1}$, where we have taken $\rho\sim 10^{15}\rm\, g\,cm^{-3}$ for the typical core density, a NS radius $R\sim 10\rm\, km$, and a toroidal field strength $B_\phi = 10^{16}B_{\phi,16}\rm\, G$. On a timescale $\xi t_{\rm A}$, the magnetic energy associated with the toroidal fields $E_{\rm B}=B_\phi^2 R^3/6$ ($B_\phi$ being the volume averaged field strength) is dissipated, giving rise to a dissipation rate of $\dot{E}_{\rm B} = E_{\rm B}/(\xi t_{\rm A})$. If the free energy associated with differential rotation is $E_{\rm dr}\sim 10^{53}\rm\, erg$, then the timescale for removing differential rotation is given by $t_{\rm diff} \sim E_{\rm dr}/\dot{E}_{\rm B}\sim (4\mathrm{\,s})\, \xi E_{\rm dr,53} E_{\rm B,50}^{-3/2}$. We see that, even for a conservative choice of $\xi\sim 1$, $E_{\rm B}\gtrsim 2.5\times 10^{50}\rm\, erg$ is required to remove the differential rotation in about $1\rm\, s$ (i.e. to \textit{increase} the rotation rate of the inner core to that of the outer regions). Such a strong magnetic field ($B_\phi\gtrsim 4\times 10^{16}\rm\, G$) is energetically possible \citep{Ciolfi2019}, but it may inevitably lead to a very large viscosity in the outer region (where the density is much lower) and to rapid spin-down of the remnant.

The consequence may be that, before the inner core is spun up, most of the energy dissipation occurs in the MRI-dominated outer region and that the majority of the angular momentum is rapidly transported to large radii $\gg 10\rm\, km$. In that case, the mass added to the core has a specific angular momentum that is smaller than the minimum angular momentum of a uniformly rotating SMNS of the same mass. Accretion of low angular momentum gas, along with neutrino cooling, will increase the central density \citep{kaplan14_thermal_pressure}, so it is possible that the remnant collapses to a black hole because the rotation energy is much less than the mass shedding limit $E_{\Omega}\ll E_{\rm \Omega,max}$ (as discussed in \S \ref{sec:slowrot}). After the collapse, the gas with larger specific angular momentum than that of the black hole, but less than that of the innermost stable orbit, will quickly plunge into the black hole. The outermost layers will then slowly accrete onto the black hole and power a GRB jet.
As we have shown in \S \ref{sec:plateaus} and \ref{sec:magboost}, even a modest reduction of $E_{\Omega}$ by a factor of 2 compared to $E_{\Omega,\rm max}$ strongly suppresses the imprints of long-lived magnetar remnants on GRB and all-sky radio data, and leads to a self-consistent understanding of these observations within a black hole central engine framework.

Finally, we note that our results supporting a black hole central engine powering short GRBs may also be extendable to long GRBs. The chief reason being the similarity in many of the observed properties between the two types of GRBs (e.g. in terms of the luminosities, the bulk Lorentz factors and the prompt GRB light curve morphology and spectral shape). Occam's razor would suggest that the simplest explanation is that they share a central engine and jet launching mechanism. An additional consideration is the total mass involved in a stellar collapse leading to a long GRB, which could apriori easily exceed $M_{\rm max}$ and lead to a proto-NS that would not remain stable for long enough to power the observed GRB.

Searches for conclusive proof for existence / nonexistence of lived NS remnants in GRBs are of significant importance to the field. If future evidence will be able to conclusively prove the existence of these objects, this will require a serious re-evaluation of our physical understanding as outlined above and undoubtedly lead to interesting astrophysical insights.

{\it Acknowledgments.} 
We thank Pawan Kumar, Brian Metzger and Kenta Hotokezaka for helpful discussions.
PB was supported by the Gordon and Betty Moore Foundation, Grant GBMF5076. WL was supported by the David and Ellen Lee Fellowship at Caltech and Lyman Spitzer, Jr. Fellowship at Princeton University.

\appendix
\section{Distribution of ejecta masses and extracted energies}
\label{app:MejEext}
\begin{figure}
\centering
\includegraphics[width = 0.4\textwidth]{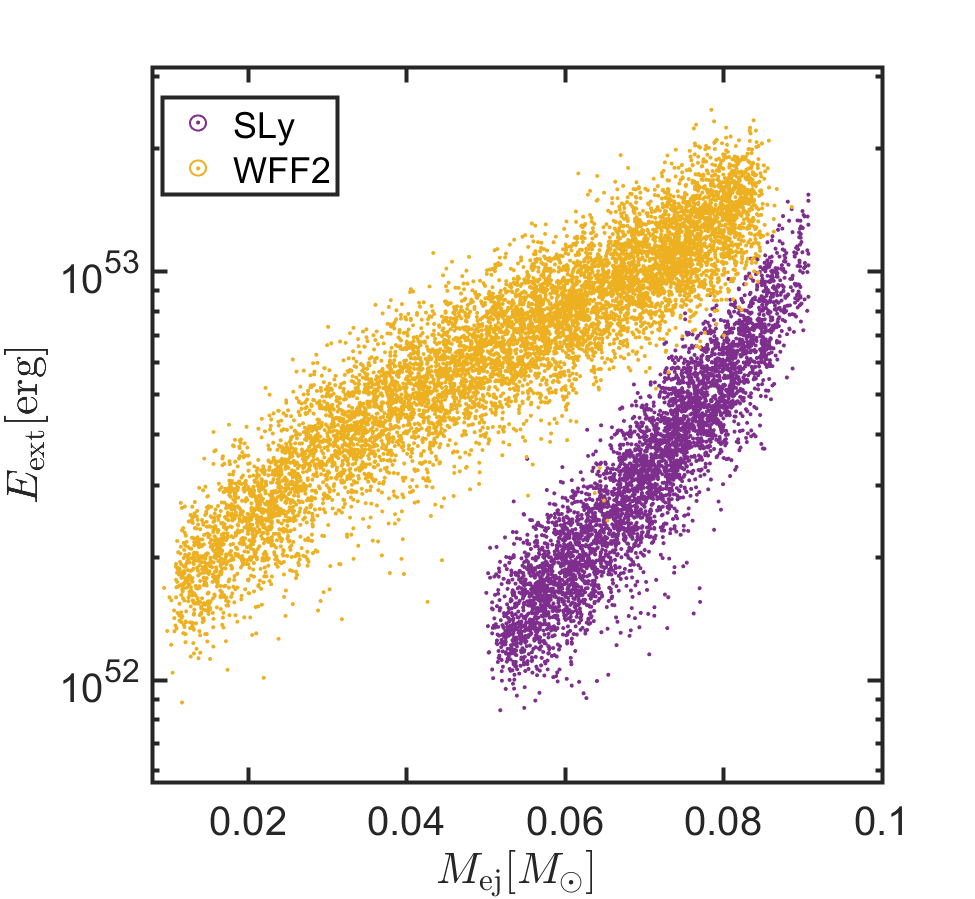}
\caption{Extracted energies as a function for ejecta masses for the two EoSs used in this paper assuming the remnant is initially rotating at the mass shedding limit. The data here is shown for systems that produce SMNS or stable NSs. Lower ejecta masses (leading to HMNS or prompt collapse) are possible, but are not associated with a significant extracted energy, and therefore not seen in this figure.}
\label{fig:EemMej}
\end{figure}

The distribution of extracted energies and ejecta masses for the EoSs used in this work are presented in figure \ref{fig:EemMej}.

\end{document}